\documentclass[graybox, envcountchap]{svmult}

\usepackage{mathptmx}        % selects Times Roman as basic font
\usepackage{amsmath}
\usepackage{amssymb}
\usepackage{color}
\usepackage{helvet}          % selects Helvetica as sans-serif font
\usepackage{courier}         % selects Courier as typewriter font
\usepackage{dirtree}
\usepackage{makeidx}        % allows index generation
\usepackage{graphicx}        % standard LaTeX graphics tool
\usepackage{subfig}

\usepackage{multicol}        % used for the two-column index
\usepackage[bottom]{footmisc}% places footnotes at page bottom

\usepackage [utopia]{mathdesign}
\usepackage [OMLmathrm,OMLmathbf,sfdefault=fav,scaled=0.875]{isomath}

\usepackage{hyperref}        %for hyperlinks
\hypersetup{colorlinks=true,urlcolor=blue}

\usepackage[misc]{ifsym}

\usepackage{CJK}

\makeindex             % used for the subject index
\def\Ledd{L_{\rm{Edd}}}
\def\Medd{\dot{M}_{\rm{Edd}}}
\def\msun{M_{\odot}}

\begin{document}
%\begin{CJK*}{UTF8}{gbsn}

%%%%%%%%%%%%%%%%%%%%%%%%%%%%%%%%%%%%%%%%%%%%%%%%%%%%%%%%%%%%%%%%%

\title{Numerical Simulations of Super-Eddington Accretion Flows}
%\author{Yan-Fei Jiang (姜燕飞) and Lixin Dai (戴丽心)}
%\institute{First Author (\Letter) \at Yan-Fei Jiang (姜燕飞), %Center for Computational Astrophysics, Flatiron Institute, New %York, NY 10010, USA, \email{yjiang@flatironinstitute.org}
%\and Second Author (\Letter)\at Lixin Dai (戴丽心), Department of %Physics, The University of Hong Kong, Pokfulam Road, Hong Kong %\email{lixindai@hku.hk}
%}

\author{Yan-Fei Jiang and Lixin Dai }
\institute{First Author (\Letter) \at Yan-Fei Jiang, Center for Computational Astrophysics, Flatiron Institute, New York, NY 10010, USA, \email{yjiang@flatironinstitute.org}
\and Second Author (\Letter)\at Lixin Dai, Department of Physics, The University of Hong Kong, Pokfulam Road, Hong Kong \email{lixindai@hku.hk}
}

\maketitle
%\end{CJK*}

\abstract{In this chapter, we summarize recent progress on the properties of accretion disks when the accretion rate exceeds the so-called Eddington limit based on multi-dimensional radiation magnetohydrodynamic simulations. We first summarize the classical models that are used to describe the accretion disks in the super-Eddington regime with an emphasis on the key uncertainties in these models. Then we show that radiation-driven outflows are ubiquitously found by numerical simulations of super-Eddington accretion disks. Some key physical processes on energy transport inside the disk are also identified by numerical simulations. Radiative and mechanical output as a function of mass accretion rates, black hole mass, spin, and magnetic field topology are summarized. Applications of super-Eddington accretion disks to different astrophysical systems, particularly tidal disruption events, are also discussed. }

%%%%%%%%%%%%%%%%%%%%%%%%%%%%%%%%%%%%%%%%%%%%%%%%%%%%%%%%%

\section{Why is Super-Eddington Accretion not only interesting but also relevant for Astrophysics?}
\label{sec:1}
Eddington luminosity $\Ledd$ \cite{Eddington1926} is defined as the luminosity when the radiation force balances the gravity for spherical symmetric system $\Ledd \equiv 4\pi GM c/\kappa$, where $G$ is the gravitational constant, $c$ is the speed of light, $M$ is the mass of the central object while $\kappa$ is the opacity. This is a well defined quantity for a given opacity $\kappa$ only when radiation force (via radiation flux) and gravity both change with radius as $r^{-2}$. If the luminosity is generated via the accretion process, the corresponding mass accretion rate is the Eddington accretion rate $\Medd\equiv \Ledd/(\eta c^2)$, where $\eta$ is the radiative efficiency and $c$ is the speed of light. Typically, electron scattering is used to estimate $\Ledd$ for the fully ionized gas, which gives a value $\Ledd=1.47\times 10^{39} \left(M/10\msun\right)\left(0.34\ \text{cm}^{2}\ \text{g}^{-1}/\kappa\right) \ \text{erg/s}$, where $\msun$ is the solar mass. For spherical symmetric systems, as in the case of stars where the concept of Eddington limit was originally discussed, it is generally expected that the large radiation force will significantly modify or even stop the accretion flow. Some models have proposed that with sufficiently high accretion rates (e.g., $>5000\Ledd/c^2$), photons are fully coupled to the gas and radiation pressure will behave like the thermal pressure with a different adiabatic index. Therefore, accretion can proceed in a similar way as described by the Bondi solution \cite{Inayoshi:2015pox,Hu:2022qnm}. However, the issue is what will happen beyond the so-called photon trapping radius
\cite{Begelman:1979}, where photon diffusion speed becomes larger than the inflow speed as density and optical depth will drop with increasing radius. Variations of opacity with temperature will also make the situation more complicated. 
If the radiation field is fully thermalized, the effective temperature at $1000$ gravitational radii $r_g$ for a given Eddington luminosity will vary with black hole mass as $T=1.52\times 10^3 \left(M/10^8\msun\right)^{-1/4} \text{K}$. 
Therefore, the gas may actually not be fully ionized for supermassive black holes with an accretion rate significantly above the Eddington limit around the photon trapping radius. A more realistic and common situation is that accretion will happen via the disk configuration due to finite angular momentum. Once the spherical symmetry is broken, it is possible that the mass inflow and escape photons can happen in different directions\cite{Ohsuga:2005bh,Krumholz:2009mi}.

Super-Eddington accretions are believed to happen in many astrophysical systems. The one with the most extensive evidence is ultraluminous X-ray (ULX) sources. These are non-nuclear, point-like objects with an apparent isotropic X-ray luminosity at least once observed with a value larger than $10^{39}$ erg s$^{-1}$, which is the Eddington luminosity of typical stellar mass black holes \cite{Feng:2011pc,Kaaret:2017tcn}. Coherent pulsations are detected in some ULXs \cite{Bachetti:2014qsa,Furst:2016mgk,Israel:2016sxx,King:2023nft}, which indicate that at least some ULXs are accreting neutron stars. Although sub-Eddington accretion onto intermediate mass black holes is still possible for some ULXs (such as HLX-1) \cite{Farrell:2009uxm}, spectra of most ULXs are inconsistent with sub-Eddington accretion disks as found for many X-ray binaries. Outflows are commonly observed for ULXs. The inferred kinetic energy flux based on the properties of shock-ionized ULX nebulae is typically comparable or even larger than the radiation flux \cite{Feng:2011pc}. Photoionized nebulae are also found for at least a few cases \cite{Pakull:2002pj,Kaaret:2004vk,Kaaret:2017tcn,Lehmann:2004rg,Kaaret:2009xh}, which suggests that the observed super-Eddington luminosity in this case is comparable to the intrinsic luminosity of the system and it cannot be just caused by strong relativistic beaming as suggested by some models \cite{Koerding:2001am,King:2008ie,Lasota:2023xef}. All these observational properties of ULXs put significant constraints on models of super-Eddington accretions onto black hole and neutron star systems. 

Super-Eddington accretion has also been commonly used to explain the rapid growth of black holes in the early Universe \cite{Volonteri:2005fj,Inayoshi:2019fun}. More than 200 quasars with black hole mass $\gtrapprox 10^9\msun$ have been found at redshift $z \ge 6$ when the Universe is less than $\sim 1$ Gyr old \cite{SDSS:2001emm,Wu:2015phw,Jiang:2016eir,Banados:2017unc}. It is necessary for the black holes to accrete with a super-Eddington rate at least for a fraction of the lifetime in order to grow from the stellar mass black hole range if the mass of the black hole seeds is $\lesssim 10^4\msun$ \cite{Volonteri:2021sfo,Mayer:2018vrr}, although this can be avoided in some models if black holes can grow via mergers \cite{Tanaka:2008bv}. The biggest uncertainty for the growth of black holes via super-Eddington accretion is how the radiative and mechanical output from the accretion disk will interact with the gas supply at large distances, which may limit the duty cycle of the super-Eddington accretion phase \cite{Pezzulli:2017ikf,Massonneau:2022sye,Mayer:2018vrr}. It has also been suggested that supermassive black holes in narrow-line Seyfert-1 galaxies or some X-ray weak quasars are also going through super-Eddington accretion phases\cite{Collin:2004qe,Komossa:2006gc,Luo:2015eua,Martocchia:2017bou,Jin:2017aez,Ma:2024}. This topic has been recently reviewed by \cite{Mayer:2018vrr} and we will not repeat it here.

%\textcolor{blue}{Discuss the potential super-Eddington %accretion for transient events, such as TDEs, stellar %black holes in AGN disks. }

Moreover, super-Eddington accretion happens in various transient accretion phenomena around black holes. For example, stars venturing too close to massive black holes (MBHs) can get destroyed by the tidal force \cite{Rees:1988bf}. In such a tidal disruption event (TDE), roughly half of the stellar debris stays bound to the black hole and is then supplied to the black hole in the timescale of years. The peak debris mass ``fallback rate'' typically exceeds the black hole Eddington accretion rate by 1-2 orders of magnitude \cite{Evans:1989qe, Guillochon:2012uc}. Therefore, TDEs are often considered ideal local laboratories for studying super-Eddington accretion and outflow physics around MBHs \cite{Dai:2021xcc}.

%%%%%%%%%%%%%%%%%%%%%%%%%%%%%%%%%%%%%%%%%%%%%%%%%%%%%%%%%

\section{Classical Models for Super-Eddington Accretion Disks}
\label{sec:2}

Properties of accretion disks are mainly controlled by the angular momentum 
and energy transport within the disk. In the standard $\alpha$ disk model \cite{Shakura:1972te,Lynden-Bell:1974vrx}, angular momentum transport is assumed to be caused by 
an effective viscosity, which is proportional to the total thermal pressure with a constant $\alpha$. Dissipation in the disk, which converts the gravitational potential energy of the accreted gas to the thermal energy, is entirely balanced by radiative cooling locally. The typical inflow speed of such a disk is \cite{Pringle:1981ds,Davis:2020wea}
\begin{eqnarray}
v_{\text{in}}&=&\alpha \Omega r \left(\frac{h}{r}\right)^2\nonumber\\
&=&5.4\times 10^4 \left(\frac{\alpha}{0.1}\right)\left(\frac{h}{0.01r}\right)^2\left(\frac{30r_g}{r}\right)^{1/2} \text{cm}/\text{s},
\end{eqnarray}
where $\Omega$ and $h$ are the angular velocity and the disk scale height at $r$, while $r_g$ is the gravitational radius for a $10^8\msun$ black hole. The averaged vertical photon diffusion speed is $c/\tau$, where $\tau$ is the vertical optical depth from the disk midplane to the photosphere. The assumptions made in the standard thin disk model is justified when $\tau\lesssim 10^5$ as the photon diffusion speed is faster than the inflow speed. 

When the accretion rate exceeds the Eddington limit, $h/r$ is expected to increase and the inflow speed can be comparable or even larger than the vertical photon diffusion speed as $\tau$ is also likely to be larger. Therefore, some of the dissipated energy will be advected towards the black hole with the accretion flow before the photons have time to diffuse out of the system. This is the most important modification introduced by the slim disk model in order to describe the structures of super-Eddington accretion disks. A stationary, vertically integrated slim accretion disk model was constructed by \cite{Abramowicz:1988sp}. It represents a natural extension of the standard thin disk model by including the pressure gradient and advective energy transport along the radial direction. When these terms become significant for $\dot{M} > \Medd$, the surface density $\Sigma$ starts to increase with increasing mass accretion rate in contrast to the $\dot{M}-\Sigma$ relation in the radiation pressure dominated thin disk solution. This suggests that the super-Eddington disk model will not subject to the viscous instability as in the thin disk case \cite{Lightman:1974sm}. The additional energy transport mechanism due to radial advection also helps avoid the thermal instability \cite{Shakura:1976xk,Piran:1978,Jiang:2013aoa}. When the accretion rate exceeds $\Medd$, the total radiative luminosity will also become larger than $\Ledd$. But then it will only increase with mass accretion rate logarithmically instead of linearly for $\dot{M} > \Medd$. The ratio between disk scale height and radius $h/r$ can increase to $\sim 0.7$ when $\dot{M}=10\Medd$. 

Various modifications to the original slim disk model have been proposed to account for different physical processes that may be important for accretion disks in the super-Eddington regime. It has been suggested that large density inhomogeneities on scales much smaller than the disk scale height can be produced in the radiation pressure dominated accretion disks due to photon bubble instability \cite{Begelman:2002rr}. This may allow photons to leave the disk much quicker compared with the photon diffusion rate as estimated based on the vertically integrated optical depth because photons will preferentially go through low density regions as suggested in the envelope of massive stars \cite{Shaviv:1998}. Therefore super-Eddington luminosity may be achieved without blowing away most of the mass. The disk scale height will also be reduced due to the enhanced cooling. Outflows have also been expected for accretion disks with super-Eddington accretion rate since the thin disk model was originally proposed \cite{Shakura:1972te}. \cite{King:2003xf} suggested that the outflow will be optically thick and cause the photospheric radius as well as outflow speed to vary with the outflow rate as $\dot{M}_o^2$ and $\dot{M}_o^{-1}$ respectively, although the actual outflow rate $\dot{M}_o$ 
for a given mass accretion rate cannot be determined self-consistently in the model.

\section{Results of Recent Numerical Simulations}
\label{sec:3}

\subsection{Radiation Hydrodynamic Simulations of Super-Eddington Accretion Disks and the Missing Physics}

To self-consistently determine the structures of super-Eddington accretion disks, including the outflow rate and radiative efficiency, multi-dimensional numerical simulations are necessary. Early two dimensional radiation hydrodynamic simulations \cite{Eggum:1988,Okuda:2002mx,Ohsuga:2005bh} model the angular momentum transport via the $\alpha$ prescription as done in the standard thin disk model. The energy and momentum exchanges between photons and gas, as well as the transport of photons, are modeled using the flux-limited diffusion approach.  These calculations confirm that super-Eddington accretion is possible when mass accretion happens near the midplane while photons leave the disk through the polar region. Temperature gradient along the radial direction is small and the optical depth near the midplane is large. Therefore, radiation force is not significant to stop the accretion flow near the midplane. For an accretion rate of $100\Ledd/c^2$, the viscous heating rate is $10\Ledd$ while the radiative luminosity is only $3\Ledd$ with the rest of the energy gets advected inward with the accretion flow. This is the "photon trapping" effect discussed in the literature. Significant optical thick outflow is launched along the polar direction, which will impact the observational properties of these systems (see Section \ref{sec:spec}). These early simulations typically didn't run long enough to determine a reliable outflow rate. 

Since magneto-rotational instability (MRI) is now believed to be the physical mechanism that generates the turbulence responsible for the angular momentum transport in accretion disks \cite{Balbus:1991ay,Balbus:1998ja}, it is crucial to include magnetic fields in simulations of super-Eddington accretion disks. Simulations in 3D are also essential to capture the properties of turbulence correctly. 
In addition, magnetic fields are necessary to produce the photon bubbles, which can enhance the radiative efficiency as proposed in some models \cite{Begelman:2002rr}. 

Another important physical process that needs to be modeled accurately is radiation transport, as radiation pressure is much larger than gas pressure for these disks, particularly for supermassive black holes. The anisotropic transport of the photons is also crucial for the launch of outflow in the super-Eddington accretion disks. Flux limited diffusion (FLD, \cite{Levermore:1981,Turner:2001jn}) is typically adopted for its simplicity, where transport of photons is approximately treated as a diffusion process. It is a valid approximation near the midplane of the disk, where it is very optically thick. However, near the photosphere, particularly near the funnel region where the outflow is launched, the diffusion approximation may not be valid. One extension is to evolve both the zeroth and first moment of specific intensities, which are radiation energy density $E_r$ and radiation flux $\vectorsym{F}_r$ respectively, in contrast to FLD where only $E_r$ is evolved. A closure relation is needed in this case to relate the radiation pressure tensor $\tensorsym{P}_r$ with $E_r$ and $\vectorsym{F}_r$ in order to close the evolution equations. A commonly adopted approach is the \emph{M1} closure, where the ratio $\tensorsym{P}_r/E_r$ is assumed to be a function of $\vectorsym{F}_r/|\vectorsym{F}_r|$ and $|\vectorsym{F}_r|/E_r$ only to connect the free streaming limit and diffusion limit \cite{Levermore:1984,Gonzalez:2007,Sdowski:2012cx,Takahashi:2013tfa,Rosdahl:2014hia}. This closure relation is local and treats photons as a fluid with an effective equation of state, which has its own limitations when the photon mean free path becomes large. A well known example is that when two streams of photons collide in the vacuum, they will merge into one beam along the direction of the net radiation flux \cite{Rosdahl:2014hia}. 
The radiation transport equation is essentially not local in this regime as photons at a particular spatial location will depend on the distribution of the sources over a large volume. The only way to get the solutions to the transport equation accurately for all parameter regimes is to calculate the solutions for specific intensities. This approach is typically more expensive as the number of variables that need to be evolved is much larger. Specific intensities can be used to calculate the 
Eddington tensor, which is needed to close the moment equations of radiation transport \cite{Stone:1992mq,Hayes:2002vh,Jiang:2012yw,Davis:2012yv,Asahina:2020,Wibking:2022}. In this scheme, which is called Variable Eddington Tensor (VET) method in the literature, the radiation field is represented by two independent sets of variables: the moments ($E_r$ and $\vectorsym{F_r}$) and specific intensities. They are only coupled via the Eddington tensor and the moments can be inconsistent with the direct angular quadrature of the intensities. Time dependent evolution equations for specific intensities can also be solved and directly coupled to the gas without evolving separate moment equations. This approach has been successfully applied for flows in both the Newtonian limit \cite{Jiang:2014tpa,Jiang:2021,Jiang:2022ccq} and general relativity (GR) \cite{White:2023wxh}.  

\subsection{Radiation MHD Simulations with GR and M1 Radiation Transport for standard and normal evolution disks}

General properties of super-Eddington accretion disks are studied based on two or three dimensional radiation magneto-hydrodynamic (MHD) simulations using the M1 closure in GR \cite{McKinney:2013txa,Sadowski:2013gua,McKinney:2015lma,Sadowski:2015hia,Sadowski:2015ena,Utsumi:2022vcc}. The numerical artifact of M1 can cause shock-like structures in the radiation field near the rotation axis of the disk, because this scheme will merge photons emitted from different azimuthal regions of the disk to one direction near the polar axis. An additional viscosity term is added to the radiation moment equations to smooth out these artificial shock structures \cite{Sadowski:2014awa,Sadowski:2015hia}.
These simulations typically start with a rotating torus located at $15-30r_g$ initially, where $r_g$ is the gravitational radius of the black hole. 
Simulation parameters are chosen to be appropriate for stellar mass black holes, up to $10^6\msun$ black holes with the opacity set to be electron scattering plus free-free absorption. Radiation MHD simulations, unlike the MHD calculations, are not scale free and therefore one simulation cannot be used to describe the accretion flow around black holes with different masses. A single poloidal loop of magnetic field is typically used in the torus to provide the seed magnetic field, which is needed for MRI. With this setup, if a significant fraction of the torus is allowed to flow towards the black hole, enough magnetic flux will be accreted near the black hole horizon to cause the disk to enter the magnetic arrested state (the MAD state). It is found that MAD super-Eddington accretion flows around fastly-spinning black holes can have unique properties, which we will describe in Section \ref{sec:MAD}. Here in this section we focus on describing the general properties of super-Eddington accretion flows which are SANE (``standard and normal evolution")  or MAD but around slowly spinning black holes. These simulations usually can only study the very inner region of the disk due to a limited run time. 
Some analyses of these simulations only focus on the initial period of $t\lesssim 1.5\times 10^4 r_g/c$, which only corresponds to one orbital period at $178.6r_g$. Therefore, these simulations can only reach a steady state for the inner $10\sim 20r_g$.

Consistent with previous 2D hydrodynamic simulations, these simulations show that gas can flow towards the black hole with a super-Eddington accretion rate near the midplane while strong outflow driven by the photons emitted by the disk is formed near the rotation axis. Outgoing mass flux only starts to show up near $\sim 10r_g$ and can become comparable to the net mass accretion rate beyond $~20-50r_g$. The mass flux is typically calculated using the time and shell averaged radial momentum for each radius. It is important to point out that not all the outflowing mass flux should be considered as truly unbound gas and different ways to determine the outflow may result in different results. For example, if outflowing gas is firstly determined for each snapshot and then the outflow is averaged over time, the total amount of inferred outflow gas can be different from the value as determined using the time averaged total radial momentum\cite{Yuan:2012yu}. The radiative luminosity integrated over the funnel region saturates to $\approx \Ledd$ while the mass accretion rate is $\approx 100 \Ledd/\left(0.057c^2\right)$ for the zero spin case shown in \cite{Sadowski:2013gua}. When the spin value is increased to $0.9$, the radiative luminosity is $\approx \Ledd$ at $20r_g$ but it continues to increase with radius to $10\Ledd$ at $200r_g$. The outflow rate is also larger with a larger spin value. The outflow generated by the super-Eddington accretion disks is also referred as "radiative jet" by \cite{Sadowski:2015ena}, although the mass weighted outflow speed is only $0.2-0.3c$ for a wide range of mass accretion rates. This is very different from the ultra-relativistic jet generated by the Blandford-Znajek mechanism with magnetic fields \cite{Davis:2020wea}. Similar outflow properties are found by simulations of super-Eddington accretion disks performed by different groups and the terminal speed is limited by strong radiation drag \cite{Jiang:2014b,Takahashi:2014oca}. The terminal speed of the outflow decreases to a few percent of the speed of light when the viewing angle is larger than $\sim 45^{\circ}$ from the rotational axis. The total kinetic power of the outflow is $\approx \Ledd$ when the mass accretion rate is $10^3\Ledd/c^2$ for the simulation done by \cite{Takahashi:2014oca}.

The isotropic equivalent luminosity as a function of polar angle and mass accretion rate for a non-spinning black hole is found by \cite{Sadowski:2015ena} to be 
\begin{eqnarray}
    L_{\text{iso}}=4\times 10^{47}\  \exp\left({\theta/\theta_0}\right)\frac{\dot{M}}{10^3\Medd}\frac{M_{\text{BH}}}{10^6M_{\odot}} \text{erg}/\text{s}.
\end{eqnarray}
The total luminosity integrated over all the angles is 
only slightly larger than $\Ledd$. It can increase with larger black hole spins.  Although the characteristic opening angle is fitted to be $\theta_0=0.2$, escaping photons are typically more strongly beamed towards the rotation axis with larger accretion rates (Figure 6 of \cite{Sadowski:2015ena}). The dependence on black hole mass also assumes that the properties of these super-Eddington disks are scale free, which cannot be tested by these simulations.

\subsection{Simulations with Full Radiation Transport}
\begin{figure}
    \centering
    \includegraphics[width=1\linewidth]{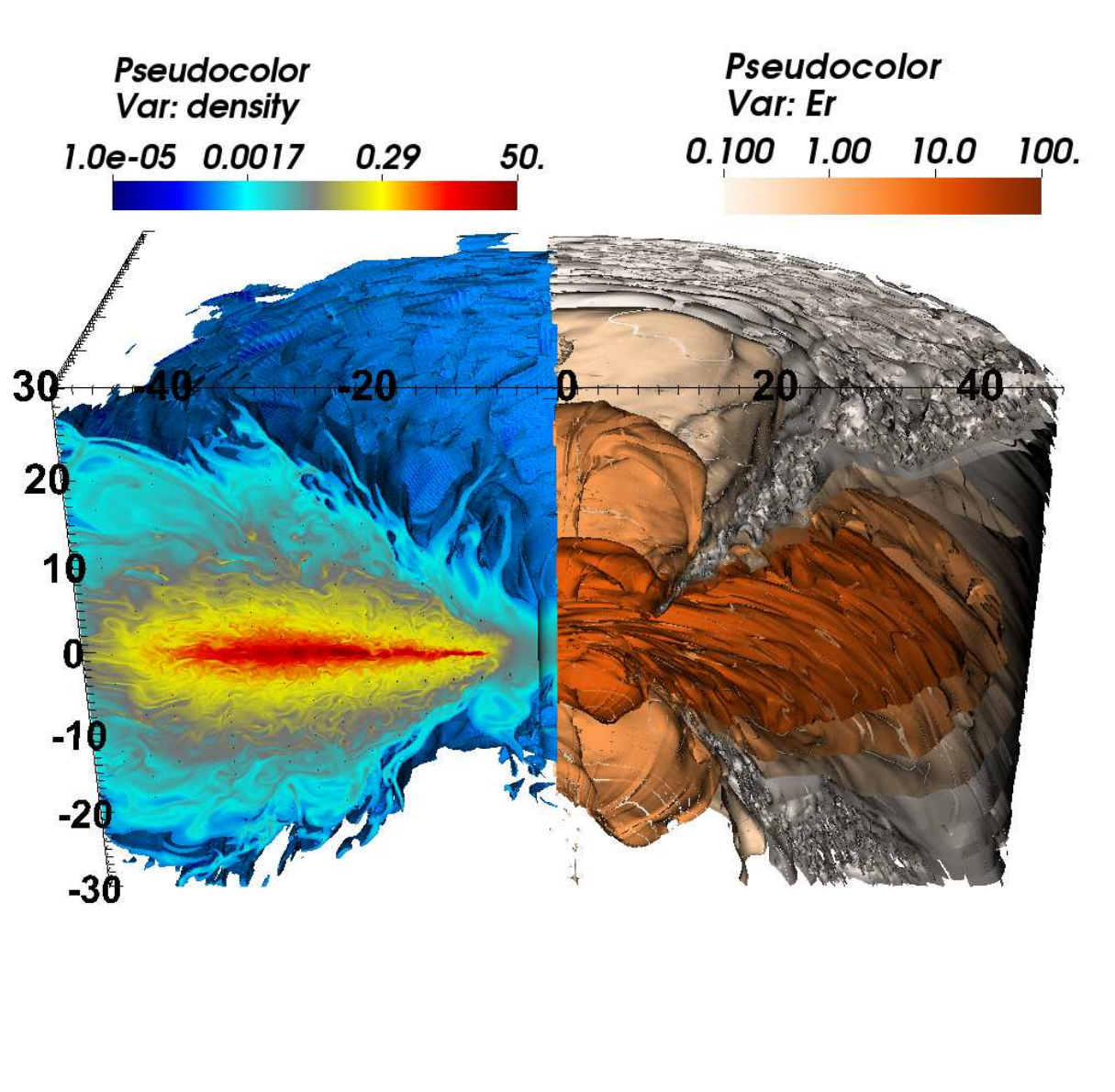}
    \caption{Typical structures of Super-Eddington accretion disks onto a stellar mass black hole. The left panel shows the density in unit of $10^{-2}$ g/cm$^{3}$ while the right panel shows the radiation energy density in unit of $a_rT_0^4$ with $T_0=10^7\text{K}$. Length scale labeled in this Figure is in unit of the Schwarzschild radius for $6.62M_{\odot}$ black hole. This Figure is adopted from \cite{Jiang:2014b}.}
    \label{fig:disk_SBH}
\end{figure}

Because radiation pressure plays an important role in determining the properties of super-Eddington accretion disks, other simulations that solve the full angular distributions of specific intensities are also available \cite{Jiang:2014b,Jiang:2017mbm,Asahina:2020,Huang:2023dux}. One approach is to solve the full time dependent evolution of specific intensities with a Newtonian flow. The GR effects around a Schwarzschild black hole are typically captured using a pseudo-Newtonian potential \cite{Paczynski:1979rz}. Using this method, \cite{Jiang:2014b} studied a super-Eddington accretion disk onto a $6.62M_{\odot}$ black hole with the accretion rate $22\Medd$. This simulation was done in cylindrical coordinates and initialized with a torus centered at $50r_g$. Typical structures for the density and radiation energy density from the disk are shown in Figure \ref{fig:disk_SBH}.  The overall properties of the disk are similar to results obtained from the simulations performed with GR and M1 closure for radiation transport qualitatively but not quantitatively. The gas is accreted near the disk midplane region while strong radiation driven outflow is launched near the rotation axis.  Total radiative luminosity emitted by the inner $40r_g$ of the disk, where a steady state is reached, is $\sim 10\Ledd$. Mass flux carried by the outflow is $29\%$ of the net mass accretion rate.

One important physical process that can enhance the cooling rate of super-Eddington accretion disks as pointed out by \cite{Jiang:2014b} is magnetic buoyancy. Density fluctuations generated by MRI turbulence in the main body of the disk are generically anti-correlated with fluctuations of magnetic pressure. Therefore, low density regions of the disk will buoyantly rise towards the disk surface, which also carries photons with them as the disk is still very optically thick. This effectively reduces the optical depth of the photons as they do not need to diffuse all the way from the midplane, but a region closer to the surface. Total radiative luminosity emitted by the inner disk that has reached a steady state is found to be $10\Ledd$ for the simulation shown in \cite{Jiang:2014b}. Similar properties are also found for super-Eddington accretion disks with different accretion rates and different black hole masses\cite{Jiang:2017mbm,Huang:2023dux}, even though these simulations are done in spherical polar coordinates not in cylindrical coordinates.  Furthermore, \cite{Jiang:2017mbm} shows that the opening angle of the funnel  region gets smaller when the accretion rate is increased. The radiative efficiency is reduced to $\sim 1\%$ when $\dot{M}$ reaches $150\Medd$.

\begin{figure}
    \centering
    \includegraphics[width=1\linewidth]{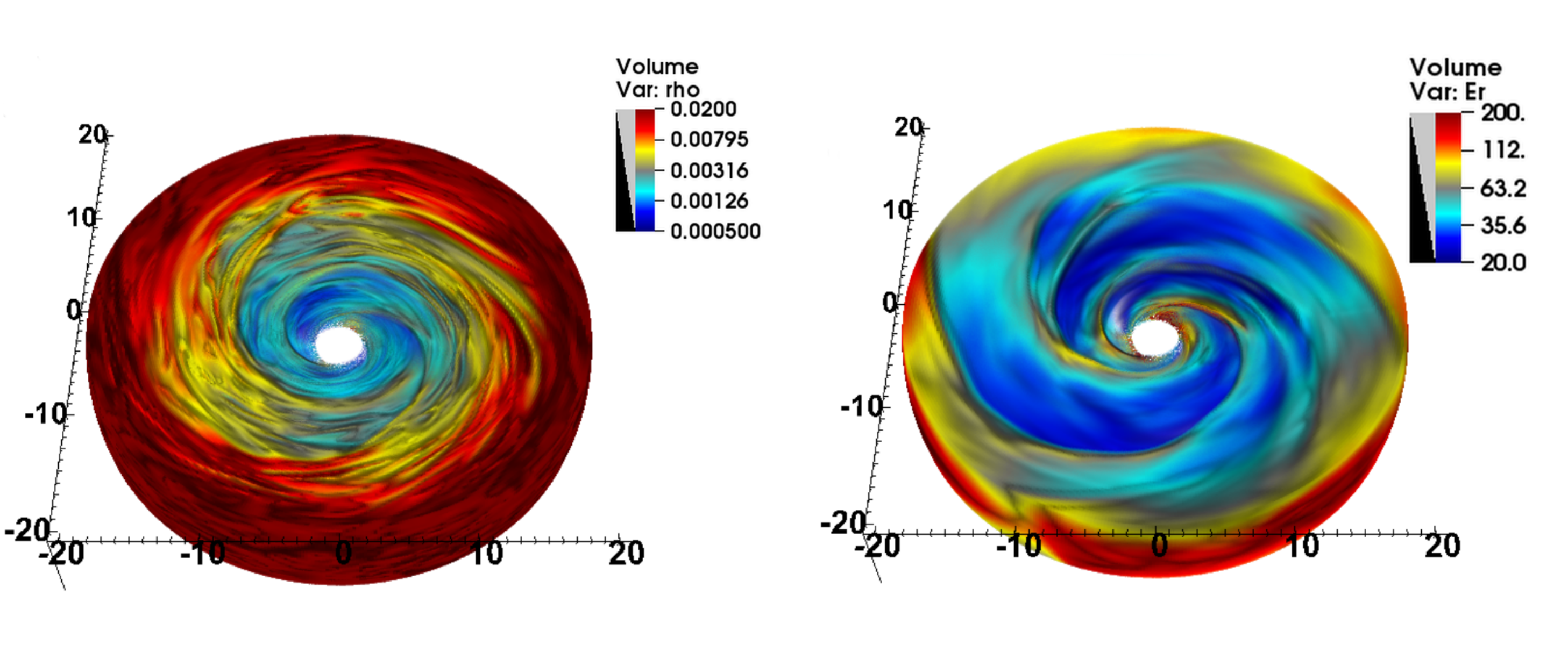}
    \caption{Spiral structures found in Super-Eddington accretion disks onto a $5\times 10^8M_{\odot}$ black holes. The left panel shows the density snapshot (in unit of $10^{-8}$ g/cm$^3$) within $40r_g$ for an accretion disk onto a $5\times 10^8M_{\odot}$ black hole with mass accretion rate $\dot{M}=25\Medd$. The right panel is the corresponding radiation energy density (in unit of $a_rT_0^4$ with $T_0=2\times 10^5$ K). Length scale labeled in this Figure is $2r_g$. This Figure is adopted from \cite{Jiang:2017mbm}.}
    \label{fig:disk_AGN}
\end{figure}

One important difference between super-Eddington accretion disks around supermassive black holes and stellar mass black holes is that the ratio between radiation pressure $P_r$ and gas pressure $P_g$ is much larger for the disks around supermassive black holes. For the two super-Eddington accretion disks onto a $5\times 10^8M_{\odot}$ black hole shown in \cite{Jiang:2017mbm}, the ratio $P_r/P_g$ can vary between $\sim 10^3$ and $\sim 10^5$ for the inner region $r<40r_g$, while the ratio is typically $10-10^3$ for the stellar mass black hole case. This difference causes the accretion flow around supermassive black holes to be much more compressible. Spiral shock structures are clearly identified in these accretion flows as shown in Figure \ref{fig:disk_AGN}. Reynolds stress caused by these spiral structures becomes comparable to or even larger than the Maxwell stress as produced by the MRI turbulence.  No significant spiral shocks are found for the stellar mass black hole case. Other properties of the accretion flow, such as radiative efficiency, mass flux and energy flux carried by the outflow for comparable accretion rates in the Eddington unit, are actually insensitive to the black hole masses. 

The numerical scheme to directly solve the equation for specific intensities with GR has been developed \cite{Davis:2019xee,White:2023wxh} and it is currently used to model the super-Eddington accretion flows. Another approach is to use the specific intensities to calculate the Eddington tensor, which is then used to close the moment equations for radiation energy density and radiation flux. Only solutions to the radiation moment equations are used to update gas energy and momentum \cite{Asahina:2020}. This can be done in the curved spacetime and it is more accurate than the M1 approach as no assumption about the closure relation is made. However, the radiation field is described by two independent sets of variables in this scheme, as the moments are not derived from the specific intensities that are used to calculate the Eddington tensor. Special care is needed to make sure they are consistent. \cite{Asahina:2020} also compared the radiation field they obtained using this scheme and the M1 closure, 
based on the flow properties as calculated by an axisymmetric general relativistic magneto-hydrodynamic (GRMHD) simulation of super-Eddington accretion disk onto a non-magnetized neutron Star as done by \cite{Takahashi:2017fix}. The solutions are very similar near the optically thick disk midplane. However, significant differences show up near the disk surface and the funnel region. In particular, the radial component of the radiation flux near the rotation axis is unphysically enhanced with the M1 closure scheme\cite{Asahina:2022}.

\subsection{Simulations of MAD Super-Eddington Accretion Flows}
\label{sec:MAD}

Strongly magnetized accretion flows show behaviors different from weakly magnetized flows, such as exhibiting non-axisymmetry in the flow structure and fast variability in the accretion rate. Also, MAD disks can potentially produce relativistic jets through magnetic processes. Several numerical works have focused on studying the properties of supper-Eddington accretion flows in the MAD limit \cite{McKinney:2015lma, Narayan:2017xup, Dai:2018jbr, Curd:2018qvy, Thomsen:2022voj, Yang:2022cpm, Ricarte:2023owr}. These works are all based on 2D or 3D general relativistic radiation magneto-hydrodynamic (GRRMHD) simulations using the M1 closure.

\begin{figure}
    \centering
    \includegraphics[width=0.8\linewidth]{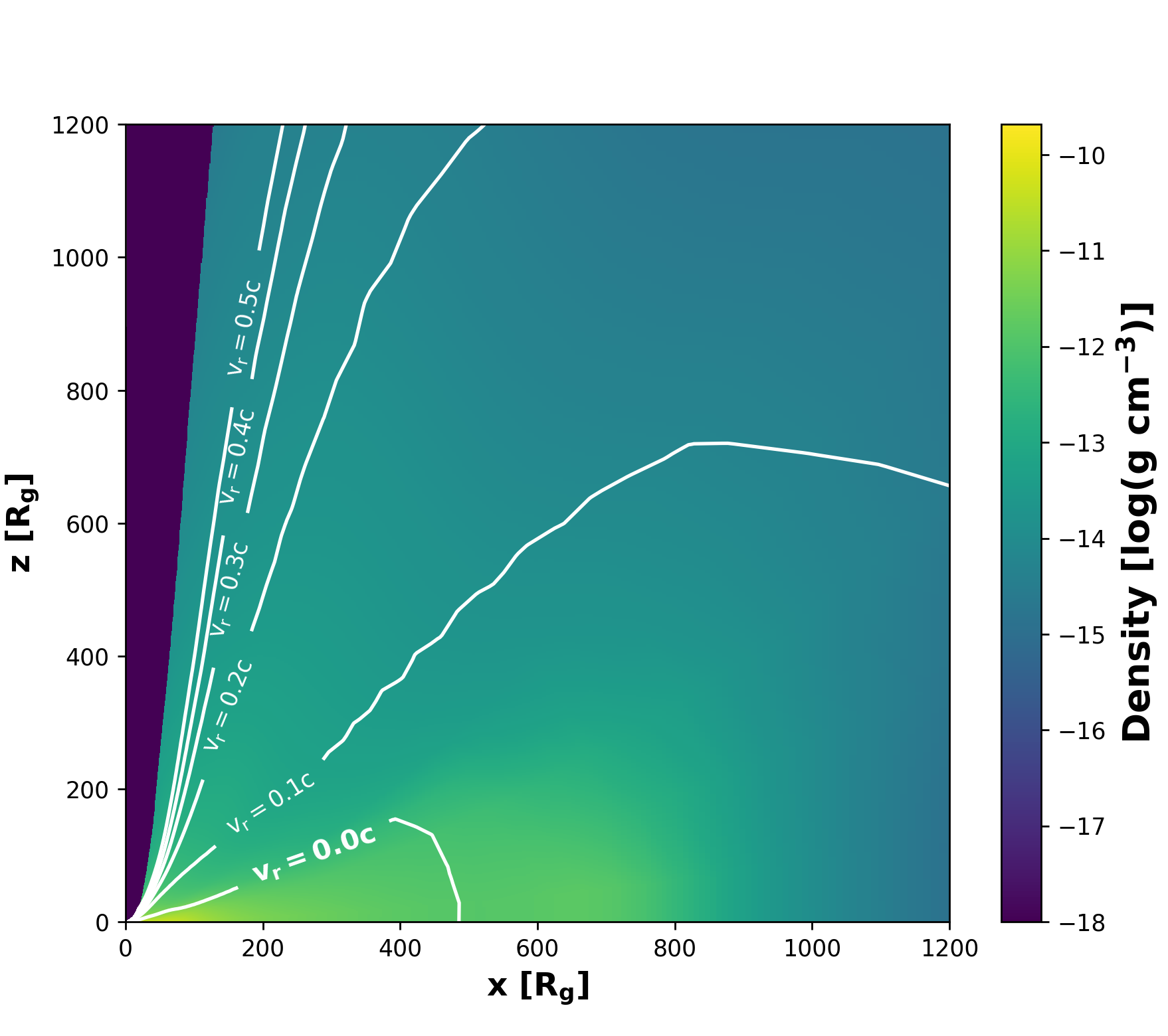}
    \caption{The time and azimuthally averaged vertical profile of a MAD super-Eddington accretion flow simulated by \cite{Dai:2018jbr}. The simulation has $M_{\rm BH}=5\times10^6 M_\odot$, $a=0.8$, and $\dot{M}_{\rm acc} \approx 15 \dot{M}_{\rm Edd}$. The gas density is shown using the background color. The white lines are contours of constant radial velocity in the wind region. A geometrically thick accretion inflow region exists around the midplane. A strong and geometrically thick wind is launched, which has an anisotropic structure. The jet region (where magnetic pressure dominates the gas pressure) is marked using a dark blue colour. This figure is adapted from \cite{Dai:2021xcc}.}
    \label{fig:disk_TDE}
\end{figure}

Winds are launched from MAD super-Eddington disks not only because of radiation pressure but also magnetic pressure. Therefore, more massive and faster winds can be produced from MAD super-Eddington disks compared to the SANE disks.  We illustrate the general accretion flow structure in Fig.~\ref{fig:disk_TDE} by plotting the accretion flow in the final quasi-equilibrium state simulated by \cite{Dai:2018jbr}. One can see that both the accretion disk and the wind are geometrically thick. Furthermore, the wind has a highly anisotropic structure in both density and velocity. ``Ultra-fast outflows'' (UFOs) with speeds faster than $0.1c$ are seen surrounding the polar region, while slower and denser winds reside between the UFOs and the disk.  Lastly, as this specific simulation involves a MAD flow around a fastly spinning black hole, a relativistic jet is launched mainly magnetically by the Blandford-Znajek mechanism \cite{Blandford:1977ds}.

The magnetized, optically thick wind launched from MAD disks helps carry radiation to an outer radius, where the wind becomes optically thin and therefore releases the radiation. When a relativistic jet is produced, the jet pushes away the wind and drills out an optically thin funnel, which allows more photons to leak out. Putting together all these factors, MAD disks around fastly-spinning black holes can reach maximum radiative efficiency and such disks can even be as radiatively efficient as thin disks.

MAD super-Eddington disks around fastly-spinning black holes also have the unique property that the disk energy efficiency generally increases with the Eddington ratio, which is opposite to the behavior of SANE disks or MAD disks around black holes of low spins. \cite{Ricarte:2023owr} further investigated the jets produced from MAD super-Eddington flows. They showed that the jets produced from such super-Eddington disks are as powerful as the jets from nonradiative MAD disks. They also gave fitting functions for how the jet power and efficiency vary with black hole spin and Eddington ratio for such disks.

\subsection{Simulations of Super-Eddington Accretion Flows in Tidal Disruption Events}
\label{sec:TDEdisk}

A series of numerical simulations have been carried out to investigate super-Eddington accretion flows formed in TDEs \cite{Shiokawa:2015iia, Sadowski:2015jor, Dai:2018jbr, Curd:2018qvy, Curd:2021zia, Thomsen:2022voj, Andalman:2020tjr, Curd:2022eoo, Bu:2022ltp, Bu:2023vjl}. One typical approach is to set up the initial disk conditions by assuming that circularized, aligned disks have already formed in TDEs, as in the simulations conducted by \cite{Dai:2018jbr, Curd:2018qvy, Thomsen:2022voj}. These simulations place the disks around supermassive black holes with masses around $10^6 M_\odot$ which are optimal for disrupting stars. One distinct property of TDE disks is that they are supposed to carry small angular momenta, since the stars are disrupted very close to the black hole. These simulations therefore either truncate the initial torus at an appropriate radius or adopt specific torus models to ensure the initial disk angular momenta is within the limit. In addition, these three works all employ 3D GRRMHD simulations with the M1 closure. The basic structure of the simulated super-Eddington accretion flows is consistent with previous simulations of super-Eddington accretion flows starting with a torus set up.

\begin{figure}
    \centering
    \includegraphics[width=0.8\linewidth]{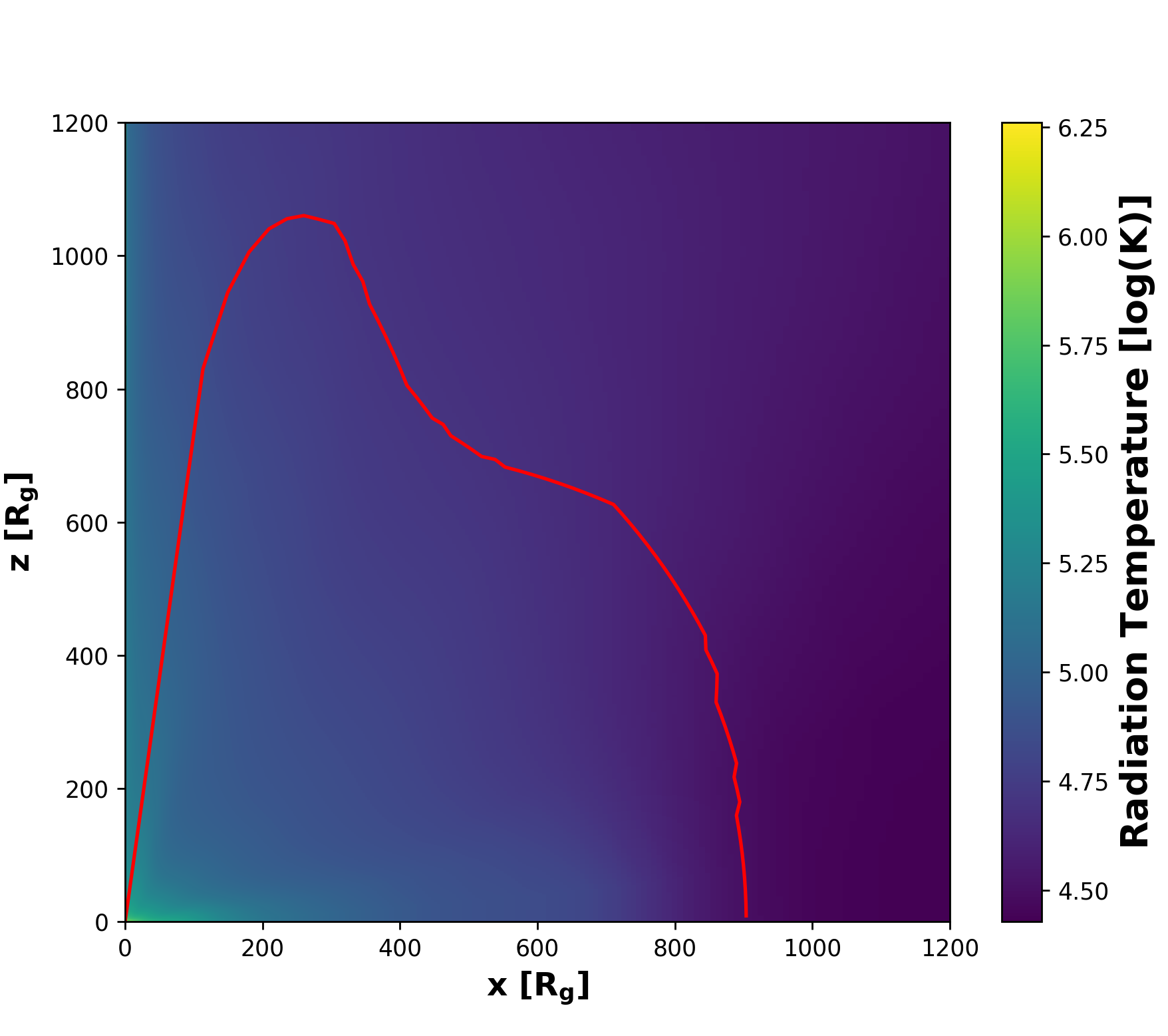}
    \caption{The time and azimuthally averaged vertical profile of the same TDE super-Eddington accretion flow as in Fig. \ref{fig:disk_TDE}. The radiation temperature is shown using the background color. The radial electron scattering photosphere with $\tau_{\rm ES}=1$ is plotted using the red line.}
    \label{fig:Trad_TDE}
\end{figure}

The anisotropic structure of the accretion flow naturally leads to a viewing angle dependence for the disk emission. Along the disk direction, the optical depth of the accretion flow is high so the radiative flux is capped around Eddington. Along the polar direction, the wind density is low so an optically thin funnel can form, which allows the radiative flux to be super-Eddington. Furthermore, for typical Eddington ratios  in TDEs ($\dot{M}_{\rm acc}\approx {\rm few} \times (1-10) \dot{M}_{\rm Edd}$), the electron scattering photosphere has a size of few$\times(100-1000) R_g$ in the wind region and truncates near the pole due to the lower wind density there, as shown in Fig.~\ref{fig:Trad_TDE}. The radiation temperature of the disk and wind ranges from $10^6$~K near the black hole to $10^4$~K in the outer wind region. The gas temperature is usually much higher than the radiation temperature, except in the disk and slow winds where the gas has a very high optical depth. More details of the emission properties of these accretion flows  will be discussed later.

The detailed properties of super-Eddington accretion flows for TDEs are demonstrated to be affected by the black hole spin and accretion rate as well as the strength of the magnetic field. The dependence on the black hole spin and the magnetic field strength has been checked by \cite{Curd:2018qvy}, which are consistent with previous results \cite{Sadowski:2015hia}. For example, they show that only when the black hole spin is large and the accretion flow is MAD, can the accretion flow have a larger efficiency in converting the gas rest-mass energy into energy, and under this scenario, most of the energy escapes in the form of a relativistic jet. The dependence on the black hole accretion rate has been investigated by \cite{Thomsen:2022voj, Curd:2022eoo}, both under the MAD limit. The former work \cite{Thomsen:2022voj} conducts a few simulations that achieve different accretion rates ranging from a few to a few tens of $\dot{M}_{\rm Edd}$ in the quasi-equilibrium state. They show that the wind structure is insensitive to the change in the accretion rate within this range. However, the wind density drops fast when the accretion rate decreases, which means that the optically thin funnel effectively opens up at a later time in TDEs. As a consequence, the inner disk emissions can more easily escape. On the other hand, the latter work \cite{Curd:2022eoo} conducts simulations with the accretion rate decreasing from a super-Eddington accretion rate of a few $\dot{M}_{\rm Edd}$ to a sub-Eddington accretion rate, so the disks are dynamically evolving throughout the simulations. They show that the disk energy efficiency can increase or decrease with the accretion rate, depending on the black hole spin. For a black hole with $a=0$, the disk becomes more radiatively efficient as the accretion rate drops. However, for the case of $a=0.9$, both the radiative efficiency and the jet efficiency generally drop with the accretion rate. These results can be applied to studying TDEs, especially the ones producing relativistic jets. Also, interestingly, \cite{Curd:2022eoo} showed that only their $a=0.9$ black hole model could produce coherent variability in the light curve, which can possibly explain the quasi-periodic oscillations (QPOs) observed in a few TDEs.

Alternatively, some TDE disk simulations do not start with an already-assembled disk or torus but instead have gas injected in certain ways mimicking the TDE disk formation process \cite{Shiokawa:2015iia, Sadowski:2015jor, Curd:2021zia, Andalman:2020tjr, Bu:2022ltp, Bu:2023vjl,Huang:2024qzx}, such as injecting gas at the TDE debris stream self-crossing points with high eccentricities \cite{Dai:2015eua}. Since simulations with such initial setups are more time-consuming, they usually do not include radiation, magnetic field, or GR, and some simulations are axisymmetric. We refer the readers to the review by \cite{Bonnerot:2020jny} for the disk formation aspects and here focus on the properties of the accretion flow after the disk is more or less assembled in these simulations. These simulations show that a significant fraction of the injected gas becomes outflows, both due to the shocks and disk radiation pressure. The structure of the outflow somewhat resembles the super-Eddington winds launched from a circularized disk such as having faster speeds with $v>0.1c$ at smaller inclinations. The outer disk usually retains an elliptical configuration at least for some period after the disk assembly, yet the inner disk eventually has an angular momentum closer to Keplerian. The disks can have a super-Eddington accretion rates under such a  configuration, which makes the inner disk advection-dominated and radiatively inefficient. Furthermore, the simulation by \cite{Sadowski:2015jor} shows that the angular momentum transfer in the newborn disks is dominated by hydrodynamical turbulence instead of magnetic fields. We note that these simulations are usually run for only a few days since the disk assembly and some of them only represent non-typical TDE parameter spaces. The long-term evolution of elliptical super-Eddington TDE disks and their interaction with the still in-falling stellar debris require further studies.

\section{Observational Properties of Super-Eddington Flows}
\label{sec:spec}

One generic outcome of all the simulations of super-Eddington accretion disks is that optically thick outflow will be launched from the inner part of the accretion disk with typical speed varying from $\sim 0.1c-0.4c$. These outflows will push the photosphere away from the black hole and reduce the effective temperature of the emitted photons. Particularly for observers that do not directly see the funnel region, X-ray emission will be much weaker compared with the standard thin accretion disks. Since the opening angle of the funnel generally decreases and the outflow rate increases with increasing mass accretion rate, the disk spectrum will be more likely dominated by the emission from the outflow for systems with high accretion rates, except for face on observers. \cite{Narayan:2017xup} performed a series of 2D and 3D simulations of super-Eddington accretion disks onto a $10M_{\odot}$ black hole. They cover accretion rates from $\sim \Medd$ to $\sim 40\Medd$ with either zero spin or dimensionless spin parameter $0.9$. In order to predict the spectra for ULXs, these simulations are post-processed using a ray-tracing scheme by extrapolating the disk profiles to a much larger radius than the radial range where a steady state is achieved for these simulations. 
Spectra from the SANE models are not sensitive to the accretion rate but strongly depend on the viewing angle. For viewing angles near edge-on, the spectra are very soft with peaks located below $<1$ keV and the apparent luminosity is much smaller than the observed luminosity for ULXs. The spectra are much harder for face-on observers and the observed luminosity can be boosted because of beaming. 
These results are actually difficult to explain the super-Eddington luminosity for the supersoft ULXs \cite{Urquhart:2015abv} and isotropic illumination favored by observations of ULX bubbles. The spectra from MAD disks are less sensitive to viewing angles. For the zero spin case, the disk shows a soft spectrum peaked around $1$ keV for $\dot{M}=1.3\Medd$ and the spectrum becomes harder with a much broader peak between $1-10$ keV when $\dot{M}$ is increased to $23\Medd$. When the dimensionless spin value is $0.9$, the spectra are very hard and broadly extend from 1 keV to 100 keV for a variety of accretion rates and viewing angles. 

\cite{Takeuchi:2013eoa} performed 2D radiation hydro and MHD simulations of super-Eddington accretion disks onto black holes covering the radial range up to $1024r_g$. For the disk with a net accretion rate reaching $10\Medd$ and an outflow with mass flux $\approx \Medd$, they find that the outflow becomes clumpy at the scale $>400r_g$. Electron scattering optical depth across each filament is typically $\sim 1$. The spatial covering factor of these clumps is $\sim 0.3$. They attribute the formation of these clumps to Rayleigh-Taylor instability. Similar clumpy structures are also confirmed in 3D simulations\cite{Kobayashi:2018qfe} using the M1 closure scheme for radiation transport. If these results are robust, they may explain the rapid variability observed for some ULXs \cite{Middleton:2010fr}. These results also demonstrate that extrapolating the simulation result beyond the radial range they have simulated can be very unreliable. 

\cite{Mills:2023yrc} used Monte Carlo scheme to post process simulations of super-Eddington accretion disks onto a $6.62\Medd$ black hole as described by \cite{Huang:2023dux}. These calculations focus on the spectra from the funnel region (so the face-on case) as only the inner steady state region can be trusted without the extrapolation as used by \cite{Narayan:2017xup}. The Monte Carlo calculations rely on gas properties (particularly temperature) given by the simulations, which can be unreliable for the funnel region. This is because Compton scattering is important to determine the gas temperature there and frequency integrated radiation transport, which is the case for most radiation MHD simulations, cannot treat Compton scattering accurately. \cite{Mills:2023yrc} restarted the simulations using the multi-group approach \cite{Jiang:2022ccq}, which solves the Kompaneets equation for Compton scattering using 20 frequency groups. Gas temperature in the funnel region typically gets hotter compared with the original solution obtained by the frequency integrated simulations. Using the improved gas properties, \cite{Mills:2023yrc} obtained spectra that can fit the observed hard X-rays (photon frequency $\gtrsim 3$ keV) of the ULX NGC 1313 X-1 very well for both total luminosity and spectrum shape.

For the case of SMBHs/TDEs, \cite{Dai:2018jbr} and \cite{Thomsen:2022voj} post-processed their 3D simulations of TDE super-Eddington accretion disks as described in Section \ref{sec:TDEdisk} using the Monte Carlo radiative transfer code Sedona \cite{Kasen:2006ce}. Sedona evolves non-local thermodynamic
equilibrium equations and can capture the electron scattering, thermal Comptonization, and the free-free, bound-free and bound-bound processes. The latter two processes are shown to be important for accurately computing the emissions from super-Eddington accretion flows around supermassive black holes,  especially in the regions where the gas temperature is in the range of $10^4-10^5$ K. \cite{Dai:2018jbr} and \cite{Thomsen:2022voj} grouped the simulations into different inclination bins and then conducted 1D radiative transfer calculations for each bin. They showed that in the low-density fast wind regions near the pole, photons go through multiple scatterings in
the expanding outflow and  lose energy through adiabatic reprocessing. In the high-density slow wind or disk region, the X-ray photons are mainly absorbed by photoionization and then re-emitted as UV/optical photons due to recombination, free-free emission, and line emission. Notably, a viewing-angle dependent unification model for TDEs is proposed based on the simulation results, which states that X-ray TDEs are observed at low inclination angles and optical TDEs are viewed at high inclination angles upon super-Eddington accretion flows formed in TDEs.

\cite{Curd:2018qvy} also post-processed their simulated TDE super-Eddington accretion flows, using the code HEROIC \cite{Zhu:2015zca, Narayan:2015cua}. Compared to the Sedona simulations mentioned above, HEROIC is a 3D post-processor but it uses a simpler model for the frequency-dependent opacity of atomic processes. In particular, HEROIC does not fully capture bound-free processes and does not include bound-bound processes. As a result of all these factors, the spectra obtained by \cite{Curd:2018qvy} carry less viewing-angle dependence compared to  \cite{Dai:2018jbr} and \cite{Thomsen:2022voj}. Moreover, \cite{Curd:2018qvy} also computed the spectra from both TDE disks and jets, and showed that the hard X-ray emissions produced by a relativistic jet can be seen from all viewing angles.

%\textcolor{blue}{Should we show some sample ULX/TDE modelled %spectra here?}

\section{Future Directions for Simulating Super-Eddington accretion disks}

One limitation of current multi-dimensional radiation MHD simulations is that they can only cover a small portion of the inner accretion disks. This is fundamentally limited by a large scale separation in the system, as the time step of these simulations is limited by the light crossing time of each resolution element at the inner boundary while the simulation needs to run at least a few thermal timescales at the radius of interest to get some reliable results. Since low energy photons will be emitted from large radii, this means the long wavelength part of the spectra from super-Eddington accretion disks cannot be easily predicted from these simulations. Simple extrapolation of the numerical solutions to larger radii \cite{Narayan:2017xup} may not be reliable. 
With the increasing computational power, particularly the recent development of simulation tools that can take full advantage of GPUs, it will be very interesting to extend these simulations to achieve a steady state over a much larger radial range.

Another important uncertainty of current simulations for super-Eddington accretion disks is how the disk is fed from large scales, which will vary for different astrophysical systems. For simplicity, a majority of current simulations start from a rotating torus that is already close to the black hole, 
which was the same initial condition as adopted for the first global MHD simulations of accretion disks more than 20 years ago \cite{Hawley:1999xv}. The choice of initial magnetic fields can result in very different properties of the disks (such as SANE versus MAD disks). For accretion onto stellar mass black holes, the condition will depend on the properties of the companions if the disk is fed by stellar winds or Roche lobe overflow. In the case of supermassive black holes, input from studies of interstellar medium and galactic properties will be necessary as summarized by \cite{Mayer:2018vrr}. When the accretion rate exceeds $100\Medd$, it is often assumed that spherical symmetric, adiabatic flow can be a good approximation to describe super-Eddington flows onto black hole seeds in the early Universe \cite{Hu:2022qnm}. This can only be true up to a certain radius. The gas temperature will drop below $10^4$ K beyond $\sim 10^3r_g$ for $10^8M_{\odot}$ with Eddington luminosity. Opacity will drop by several orders of magnitude compared with the electron scattering value. There is just not enough opacity to keep the adiabatic condition.  Therefore,  understanding the dynamics at the large scale where the gas is fed is crucial for the long-term evolution of super-Eddington accretion. 

For TDEs, the current numerical studies can be extended to address two important, unique aspects of the disks formed in these transient events. First, it is generally expected that a TDE disk, when just formed, should retain an elliptical configuration. The long-term evolution of elliptical super-Eddington disks such as whether and how quickly they can get circularized and their emission properties throughout this course are worthy of further studies. Second, the disrupted star likely has angular momentum not along the black hole spin axis, which will lead to the formation of a misaligned super-Eddington disk. However, up to now only aligned super-Eddington accretion disks have been studied using radiation MHD simulations. It is generally expected that thick, misaligned disks should precess like solid bodies (see Chapter 11). If this is indeed the case for misaligned super-Eddington accretion disks, then we can expect that the winds launched from such precessing disks should eventually achieve a quasi-spherical geometry, and the jets, if produced, may or may not be able to drill out of the wind materials.

% extend the scale, both spatial, time and spectrum

% feeding, initial and boundary condition

%%%%%%%%%%%%%%%%%%%%%%%%%%%%%%%%%%%%%%%%%%%%%%%%%%%%%%%%%

\begin{acknowledgement}
LD acknowledges the support from the National Natural Science Foundation of China
and the Hong Kong Research Grants Council (HKU12122309, N\_HKU782/23, 17305920, 27305119, 17305523).
\end{acknowledgement}

%%%%%%%%%%%%%%%%%%%%%%%%%%%%%%%%%%%%%%%%%%%%%%%%%%%%%%%%%

%\section*{Appendix}
%\addcontentsline{toc}{section}{Appendix}

%When placed at the end of a chapter or contribution (as opposed to at the end of the book), the numbering of tables, figures, and equations in the appendix section continues on from that in the main text. Hence please \textit{do not} use the \verb|appendix| command when writing an appendix at the end of your chapter or contribution. If there is only one the appendix is designated ``Appendix'', or ``Appendix 1'', or ``Appendix 2'', etc. if there is more than one.

%%%%%%%%%%%%%%%%%%%%%%%%%%%%%%%%%%%%%%%%%%%%%%%%%%%%%%%%%

\bibliographystyle{plain}

\end{document}